\documentclass[pdflatex,sn-mathphys-num]{sn-jnl}

\usepackage{graphicx}%
\usepackage{multirow}%
\usepackage{amsmath,amssymb,amsfonts}%
\usepackage{amsthm}%
\usepackage{mathrsfs}%
\usepackage[title]{appendix}%
\usepackage{xcolor}%
\usepackage{textcomp}%
\usepackage{manyfoot}%
\usepackage{booktabs}%
\usepackage{algorithm}%
\usepackage{algorithmicx}%
\usepackage{graphicx}
\usepackage{algpseudocode}%
\usepackage{listings}%
\usepackage{subcaption}
\theoremstyle{thmstyleone}%
\theoremstyle{thmstyletwo}%

\theoremstyle{thmstylethree}%

\begin{document}

\title[Article Title]{Quantitative Global Carbon Inequality Network}

\author[1]{\fnm{Yanming} \sur{Guo}}\email{yguo0337@uni.sydney.edu.au}

\author[2]{\fnm{Charles}
\sur{Guan}}\email{charles.guan@richdataco.com}

\author[1]{\fnm{Jin} \sur{Ma}}\email{j.ma@sydney.edu.au}

\affil[1]{\orgdiv{School of Electrical \& Computer Engineering}, \orgname{University of Sydney}}

\affil[2]{\orgname{Rich Data Co}}


\abstract{International trading networks significantly influence global economic conditions and environmental outcomes. A notable imbalance between economic gains and emissions transfers persists, manifesting as carbon inequality. This study introduces a novel metric, the Ecological Economic Equality Index, integrated with complex network dynamics analysis, to quantitatively evaluate the evolving roles within the global trading network and to pinpoint inequities in trade relationships from 1995 to 2022. Utilising high spatiotemporal resolution data from the Environmentally Extended Multi-regional Input-output model, our findings reveal a widening disparity in carbon inequality and dynamic patterns. This analysis emphasises the gap in regional carbon inequality and identifies unequal trade. The study underscores that carbon inequality is a critical challenge affecting both developing and developed regions, demanding widespread attention and action.}


\maketitle

\section{Introduction}\label{sec1}
The United Nations' assignment of the Paris Agreement and Sustainable Development Goals has established global goals 1, 8, and 13, targeting poverty alleviation and the maintenance of economic growth amidst the challenges posed by climate change \cite{rogelj2016paris, fund2015sustainable}. The role of international trade is crucial in facilitating regional economic development by leveraging local competitive advantages through an effective redistribution of resources \cite{xu2020impacts, jorgenson2014economic, steinberger2012pathways}. However, the trade-climate dilemma, also known as the pollution haven effect \cite{li2020carbon, liu2016targeted}, along with the international trade of embodied emissions, which accounts for 26\% of global emissions, poses significant environmental challenges \cite{davis2010consumption, arce2016carbon}. This dual impact of global trade, coupled with the need to ensure fairness in international trade, presents a significant challenge \cite{jakob2013interpreting}. Therefore, this paper aims to investigate and assess global environmental economic carbon inequality for equitable climate change policies to promote sustainable development worldwide.

Carbon inequality within the context of global trade is defined as the uneven distribution of economic benefits and emission reduction burdens \cite{wang2022carbon}. Previous research on carbon inequality can be classified into two principal categories: populational and spatial carbon inequality. The former examines the stark disparities in carbon emissions across various income groups, revealing that about half of the global emissions are generated by the wealthiest 10\% of the worldwide population \cite{mi2020economic, chancel2022global, hubacek2017global, bruckner2022impacts}. The latter explores emissions' imbalanced distribution and driving factors across different spatial scales \cite{zhu2022unfolding, chen2015urban, du2023carbon}. It concludes that developed countries, subject to stringent carbon regulations, have outsourced some high-carbon-intensive but low-value-added industries to developing countries, thus meeting the economic growth requirements for these regions. Such outsourcing leads to a typical pattern in carbon inequality across different spatial scales \cite{padilla2013explanatory, xu2022carbon, meng2023narrowing, tian2022regional, zhang2021causes}. However, the disproportionate relationship between economic benefits and emission reduction burdens exacerbates carbon inequality. Consumption-based emissions accounting was utilised for better responsibility assignment. Developed regions have higher consumption-based emissions than their production-based emissions, revealing unequal regional positions in international trade. Fair climate policies need to be designed to mitigate these inequalities \cite{steininger2016multiple, long2018embodied, fang2020average, peters2011growth, zhang2017multi}.

In the discussion of spatial carbon inequality, three significant gaps persist. First, there is an absence of a quantitative metric to assess regional carbon inequalities in global trade. This lack leads to vague identifications, descriptions, and comparisons of regional situations. Specifically, determining the number of economic benefit units that should be exchanged for one ton of emission reduction burden to constitute a fair transaction remains unclear. Furthermore, the specific gap in carbon emissions inequality between countries needs precise measurement rather than ambiguous sequential relationships. Additionally, the evolving dynamics of these disparities over time need to be quantitatively analysed. Second, the scope of prior studies has generally been restricted to bilateral trade analyses, which offer only pairwise comparisons with limited spatial coverage, leading to a limited perspective and failing to provide a comprehensive view of the relationships among various regions \cite{duan2019economic, zhang2018unequal}. Third, poor temporal resolution hinders the discovery of dynamic patterns over time and complicates forecasting changes in carbon inequality \cite{chepeliev2023gtap, dietzenbacher2013construction}. Constraints in global data collection result in existing data being often outdated and incomplete, presented in discrete timesteps.

Answers to these standing questions are crucial in achieving fair climate policies. This paper addresses these issues by introducing the Ecological Economic Equality Index (EEEI) as a quantitative carbon inequality indicator, clarifying and examining the current state of global inequality. By leveraging complex network analysis \cite{brin1998anatomy, onnela2005intensity, saramaki2007generalizations, landherr2010critical}, this study highlights the disproportionate emission and economic flows in the global trading network, providing comprehensive spatiotemporal information to understand the evolving global carbon inequality trajectory over time. This approach broadens the local-to-global perspective by employing the open-source Environmentally Extended Multi-Regional Input-Output (EE-MRIO) dataset, ExioBase 3.8.2 \cite{stadler2018exiobase}, which covers 163 sectors across 44 countries and 5 'Rest of the World' regions spanning from 1995 to 2022. By integrating the proposed EEEI measure with the global trading network, we derive the carbon inequality network to target trades suffering from carbon inequality issues. This aims to provide quantified references on global carbon inequality for economic and political decision-makers and promote sustainable development.

\section{Results}\label{sec2}
\subsection{Spatial Carbon Inequality}
\begin{figure}[!h]
    \centering
    \begin{subfigure}[b]{0.49\textwidth}
        \includegraphics[width=\textwidth]{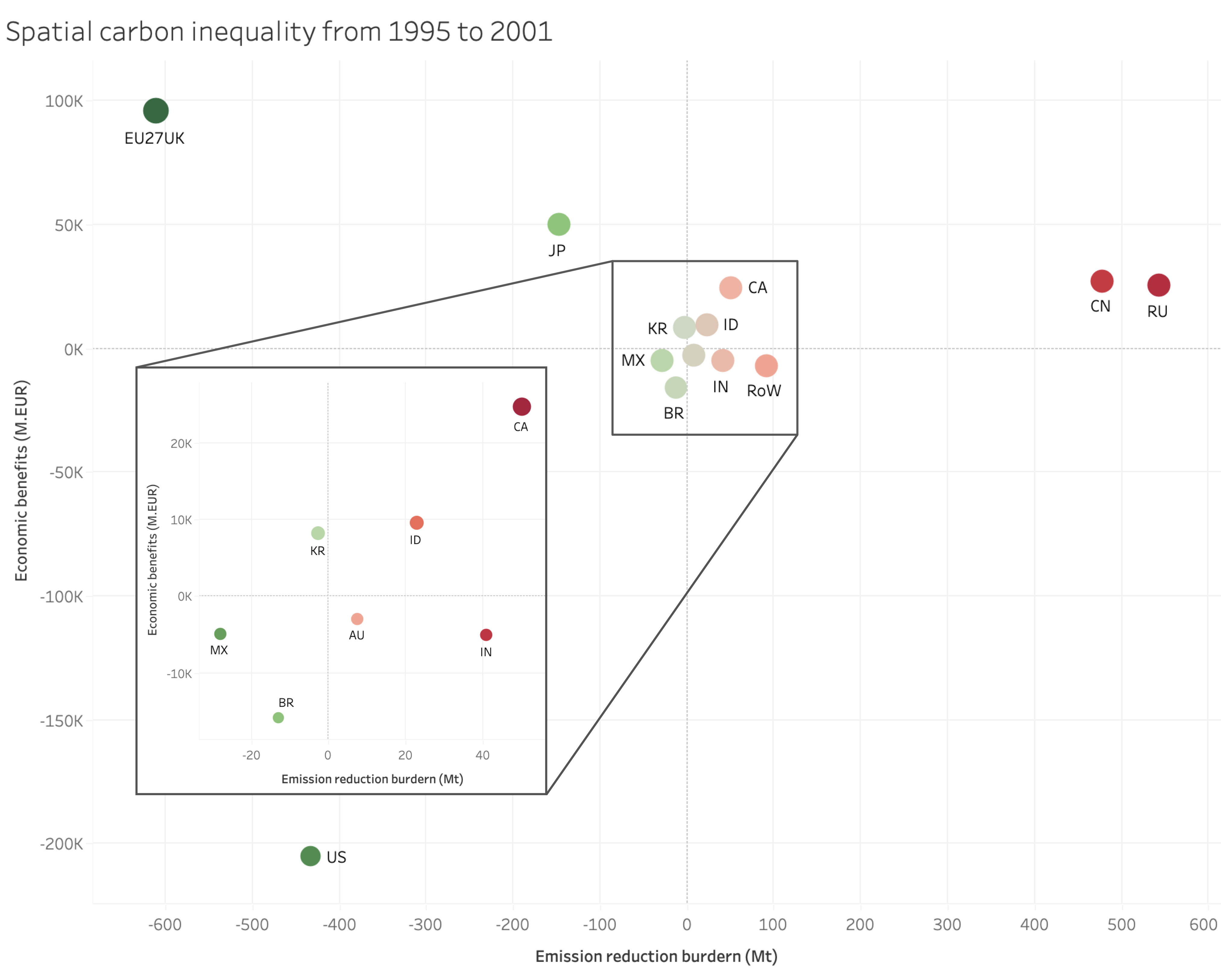}
        \caption{Spatial carbon inequality in Period 1}
        \label{fig:etd95}
    \end{subfigure}
    \begin{subfigure}[b]{0.49\textwidth}
        \includegraphics[width=\textwidth]{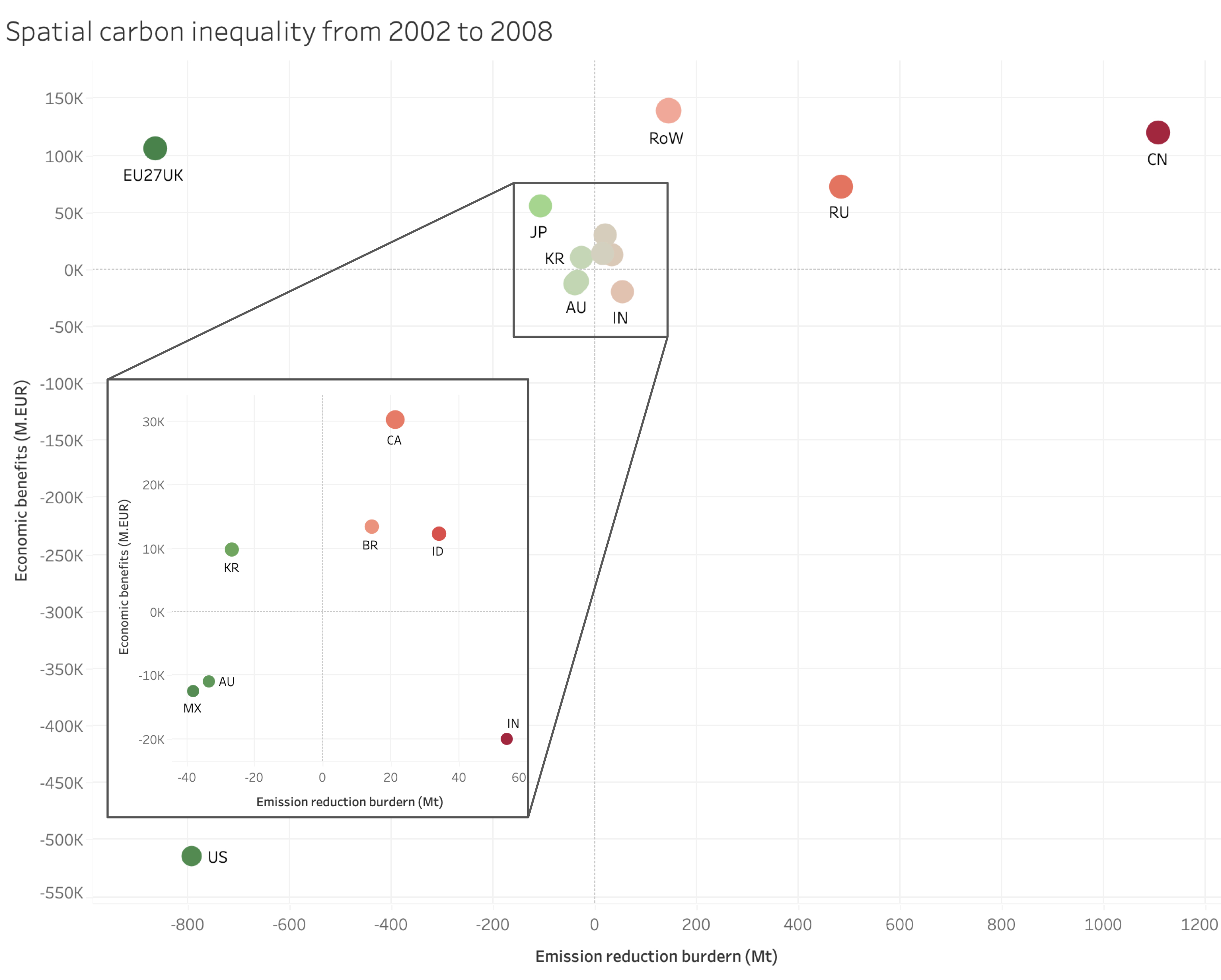}
        \caption{Spatial carbon inequality in Period 2}
        \label{fig:etd02}
    \end{subfigure}
    
    
    \begin{subfigure}[b]{0.49\textwidth}
        \includegraphics[width=\textwidth]{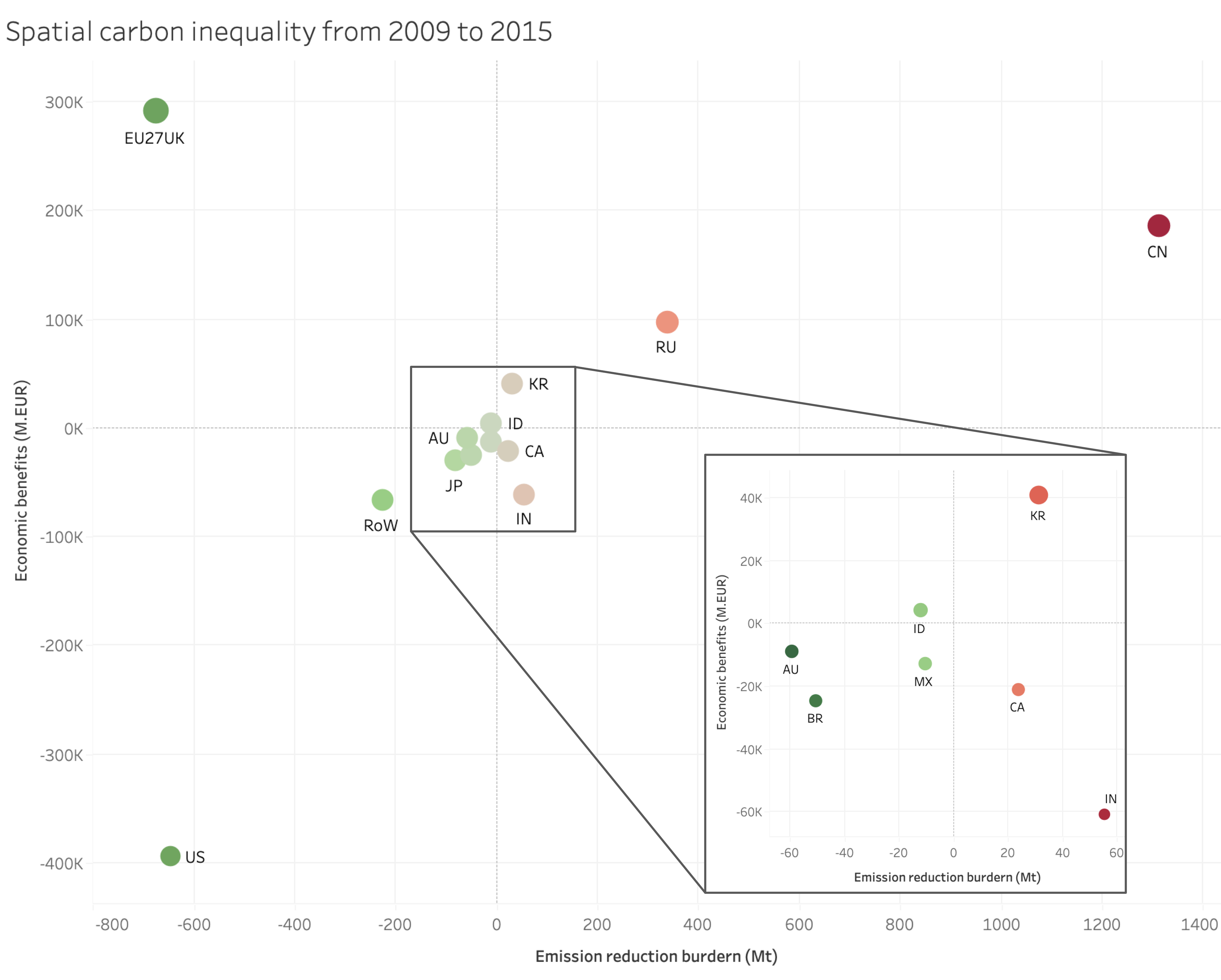}
        \caption{Spatial carbon inequality in Period 3}
        \label{fig:etd09}
    \end{subfigure}
    \begin{subfigure}[b]{0.49\textwidth}
        \includegraphics[width=\textwidth]{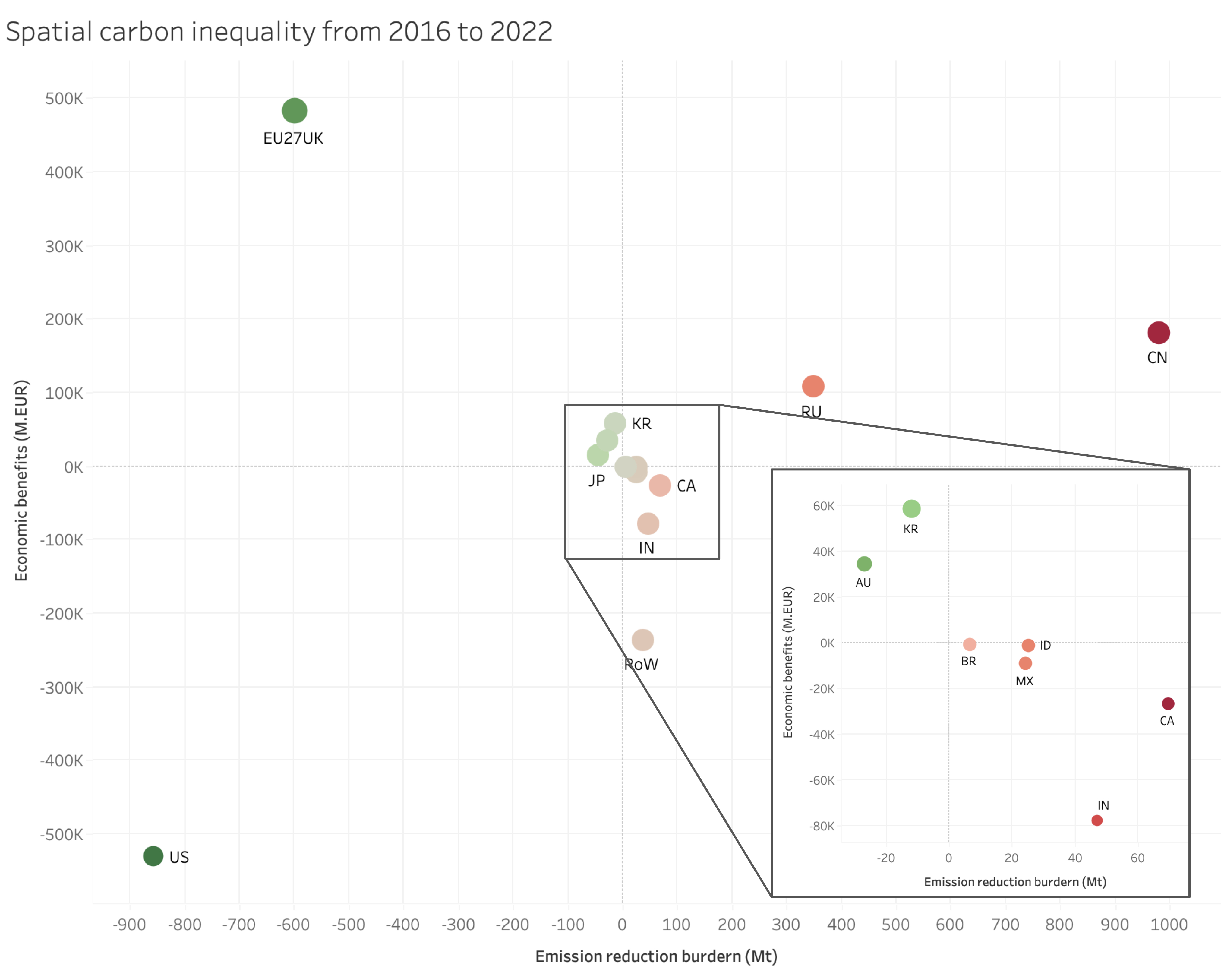}
        \caption{Spatial carbon inequality in Period 4}
        \label{fig:etd16}
    \end{subfigure}
    \caption{Regional carbon inequality is illustrated through an analysis of net emission burdens (unit in million ton, Mt) on the x-axis and net economic benefits (unit in million Euro, M.EUR) on the y-axis over four distinct periods: \textbf{1) 1995 to 2001}, \textbf{2) 2002 to 2008}, \textbf{3) 2009 to 2015}, and \textbf{4) 2016 to 2022}. The black box magnifies the regions in the middle to provide clearer visualisation. Regions such as China and Russia are positioned in the first quadrant as emission exporters with positive trade surpluses but also bear substantial emission burdens. In contrast, regions in the second quadrant, including the EU27/UK and Japan, are emission importers with lower emission burdens but substantial economic benefits.}
    \label{fig:etd}
\end{figure}

The trade-climate dilemma underscores the trade-off between economic benefits and emission reduction burdens. Spatial carbon inequality in international trade serves as evidence of the mismatch between economic gains and emission responsibilities. Figure \ref{fig:etd} illustrates the dynamic, unequal relationship among regions from 1995 to 2022, highlighting the region-specific embodied emission structures. The x-axis represents the net emissions reduction burden, while the y-axis denotes the net economic benefits, defined as the value of exports minus imports. The regions are categorised into roles within the four quadrants:
\begin{itemize}
    \item Emission exporters with a trade surplus (first quadrant, e.g. China, Russia). 

    \item Emission importer with a trade surplus (second quadrant, e.g. EU27/UK, Japan).

    \item Emission importer with a trade deficit (third quadrant, e.g. US).

    \item Emission exporter with a trade deficit (fourth quadrant, e.g. India, Canada).
\end{itemize}

These four quadrants encapsulate the roles of various regions in global trade and the carbon inequality they endure in international trade. Generally, regions in the second quadrant hold the strongest position in global trade. They not only transfer the burden of domestic carbon emission reductions to other countries through international trade but also gain substantial economic benefits. In contrast, regions as emission exporters with trade deficits in the fourth quadrant face environmental and economic pressures simultaneously. They bear significant emission reduction burdens but lack corresponding economic benefits. 

Both as countries with trade surplus over time, when comparing economic benefits, the EU27/UK consistently outperformed China. The EU27/UK secured economic gains of 96K M.EUR (period 1), 105K M.EUR (period 2), 291K M.EUR (period 3), and 483K M.EUR (period 4), while China achieved much lower economic benefits of 27K M.EUR (period 1), 119K M.EUR (period 2), 185K M.EUR (period 3), and 181K M.EUR (period 4). However, from an environmental perspective, the situations in China and the EU27/UK are completely opposite. China took substantial emission reduction burdens—478 Mt (period 1), 1109 Mt (period 2), 1313 Mt (period 3), and 981 Mt (period 4)—to obtain these economic benefits, while the EU27/UK not only achieved higher economic gains but also managed to offload significant portions of its emission reduction burdens—611 Mt (period 1), 865 Mt (period 2), 676 Mt (period 3), and 599 Mt (period 4). Furthermore, this analysis emphasises the worsening carbon inequality from 1995 to 2022. The EU27/UK’s trade surplus increased by 387K M.EUR from period 1 to 4, with little change in its emission burden. In stark contrast, China’s trade surplus only grew by 154K M.EUR, yet its emission burden surged by 503 Mt. This disparity underscores the growing carbon inequality in international trade.

\subsection{Quantitative Ecological Economic Equality Index}
\begin{figure}[!h]
    \centering
    \includegraphics[width=1\linewidth]{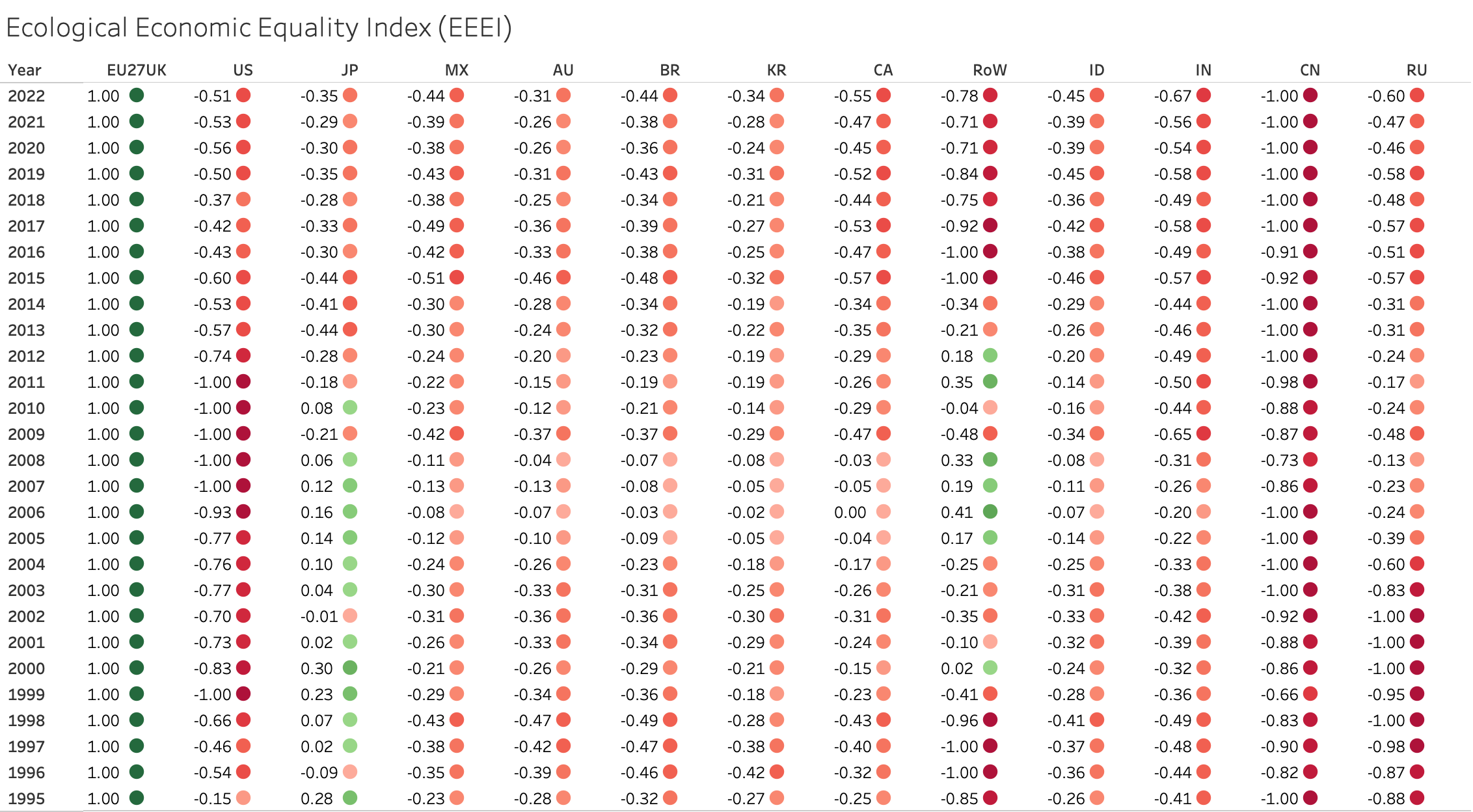}
    \caption{The dynamic trajectory of quantitative carbon inequality, encapsulated by the Ecological Economic Equality Index (EEEI), is summarised across 13 regions from 1995 to 2022. Regions with higher values indicate a favourable position within global trade dynamics. The EEEI highlights the uneven distribution of carbon inequality and observes an increasing trend, signalling a widening disparity in carbon equity across both developed and developing regions.}
    \label{fig:eeei}
\end{figure}

Although existing analyses provide insights into the roles of different countries in carbon inequality, they lack quantitative measures that can precisely describe and compare the specific circumstances of regional carbon inequality and fail to capture evolving trajectory over time, making it challenging to devise and assess effective policies for examining and improving this issue. To address these shortcomings, we propose the expandable Ecological Economic Equality Index (EEEI), which applies to any EE-MRIO framework. It considers both the economic and environmental impacts of regions in global trade, using an intuitive scale ranging from -1 to 1 to reflect regional conditions. A value of -1 indicates a disadvantaged position in international trade, while a value of 1 indicates an advantageous position. To represent the disparities between regions, we define the EEEI distance, with greater distances indicating more significant differences between regions.

The EEEI enables the rapid identification and comparison of carbon inequality across regions illustrated in Figure \ref{fig:eeei}. Consistent with previous analyses, the EU27/UK exhibits the highest trade surplus and the smallest carbon emission burden, maintaining a value of 1 from 1995 to 2022. This indicates the EU27/UK’s long-term advantageous position in the global trade. Most regions have negative values, indicating uneven distribution and varying degrees of carbon inequality compared to the EU27/UK. Countries like China, Russia, and the Rest of the World have consistently low values close to -1, signifying worse conditions compared to developed countries such as Japan, Australia, and Canada. Even Japan has a significant disparity in carbon inequality compared to the EU27/UK. While previous research has primarily focused on developing countries as the primary victims of carbon inequality, it has often overlooked the significant carbon inequality faced by some developed countries. For instance, the US experienced severe carbon inequality from 2006 to 2011, with a value of -1. Moreover, the EEEI distance effectively illustrates the increasing trend of carbon inequality. From 1995 to 2022, most regions have shown a growing distance from the EU27/UK. For example, the EEEI distance of Japan increased from 0.72 to 1.35, the US from 1.15 to 1.51, Mexico from 1.23 to 1.44, Australia from 1.28 to 1.31, Brazil from 1.32 to 1.44, Korea from 1.27 to 1.34, Canada from 1.25 to 1.55, Indonesia from 1.26 to 1.45, and India from 1.41 to 1.67.

From this analysis, we draw three important conclusions. First, carbon inequality is not evenly distributed, with most countries suffering more compared to the EU27/UK. Second, we highlight that even developed countries are at risk of carbon inequality, making it a global issue that requires attention from all regions. Thirdly, we observe that the EEEI distance for most regions is increasing relative to the EU27/UK, indicating a widening gap in carbon inequality. This trend underscores the urgent need for climate policies to mitigate these inequalities.

\subsection{Dynamic Pattern of Global Trading Network}
\begin{figure}
    \centering
    \begin{subfigure}[b]{0.49\textwidth}
        \includegraphics[width=\textwidth]{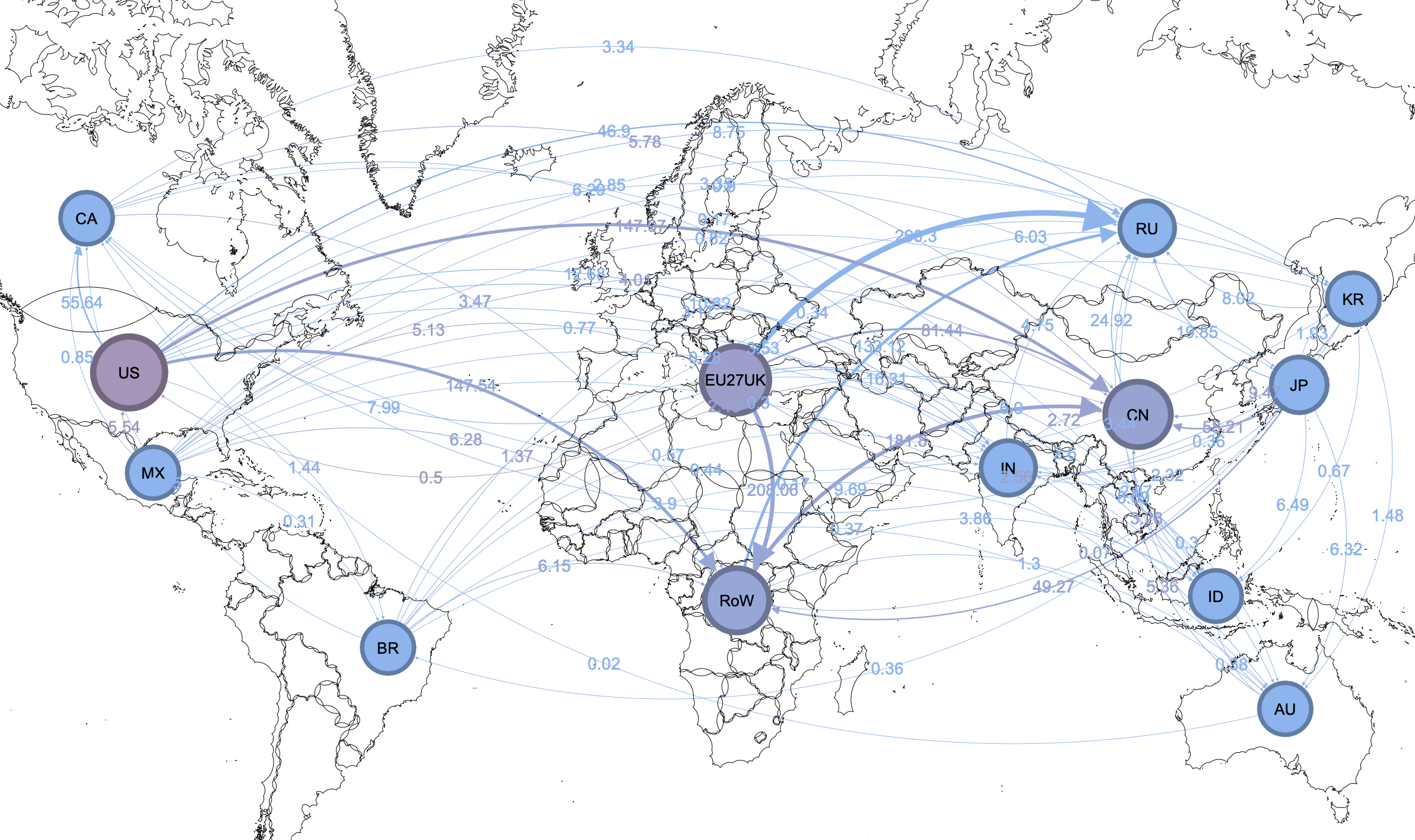}
        \caption{Emission network in period 1}
        \label{fig:en95}
    \end{subfigure}
    \begin{subfigure}[b]{0.49\textwidth}
        \includegraphics[width=\textwidth]{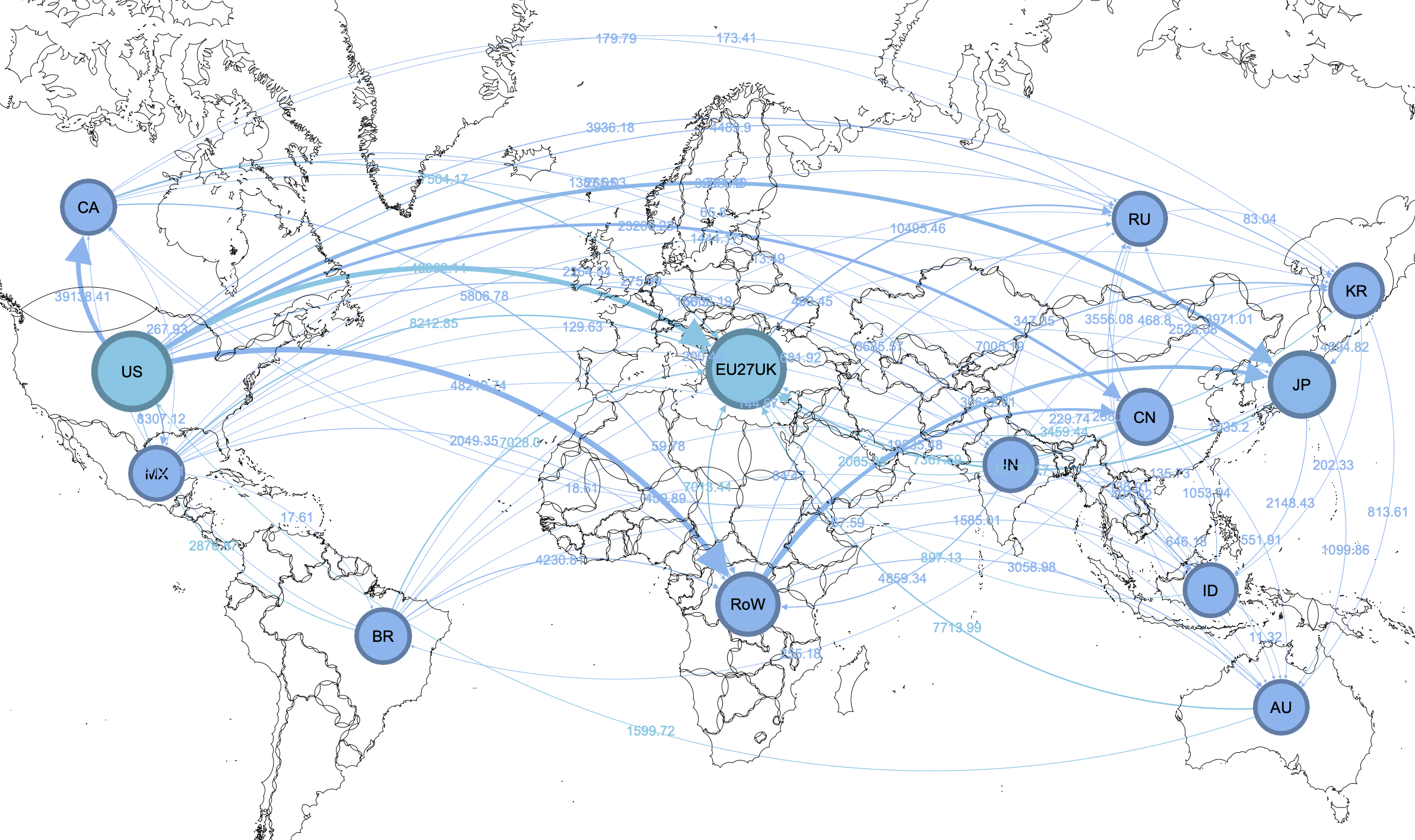}
        \caption{Value-added network in period 1}
        \label{fig:vn95}
    \end{subfigure}

    \begin{subfigure}[b]{0.49\textwidth}
        \includegraphics[width=\textwidth]{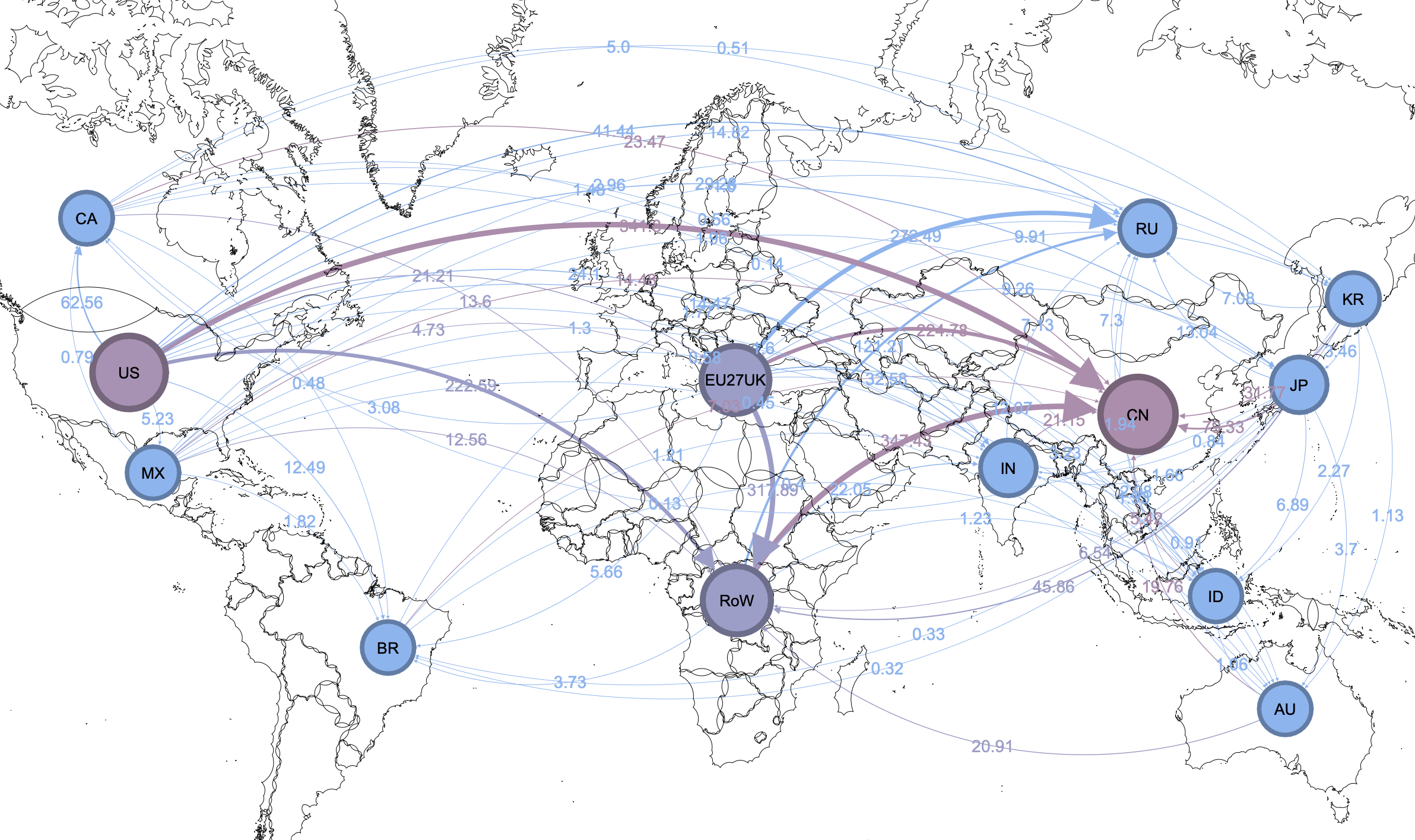}
        \caption{Emission network in period 2}
        \label{fig:en02}
    \end{subfigure}
    \begin{subfigure}[b]{0.49\textwidth}
        \includegraphics[width=\textwidth]{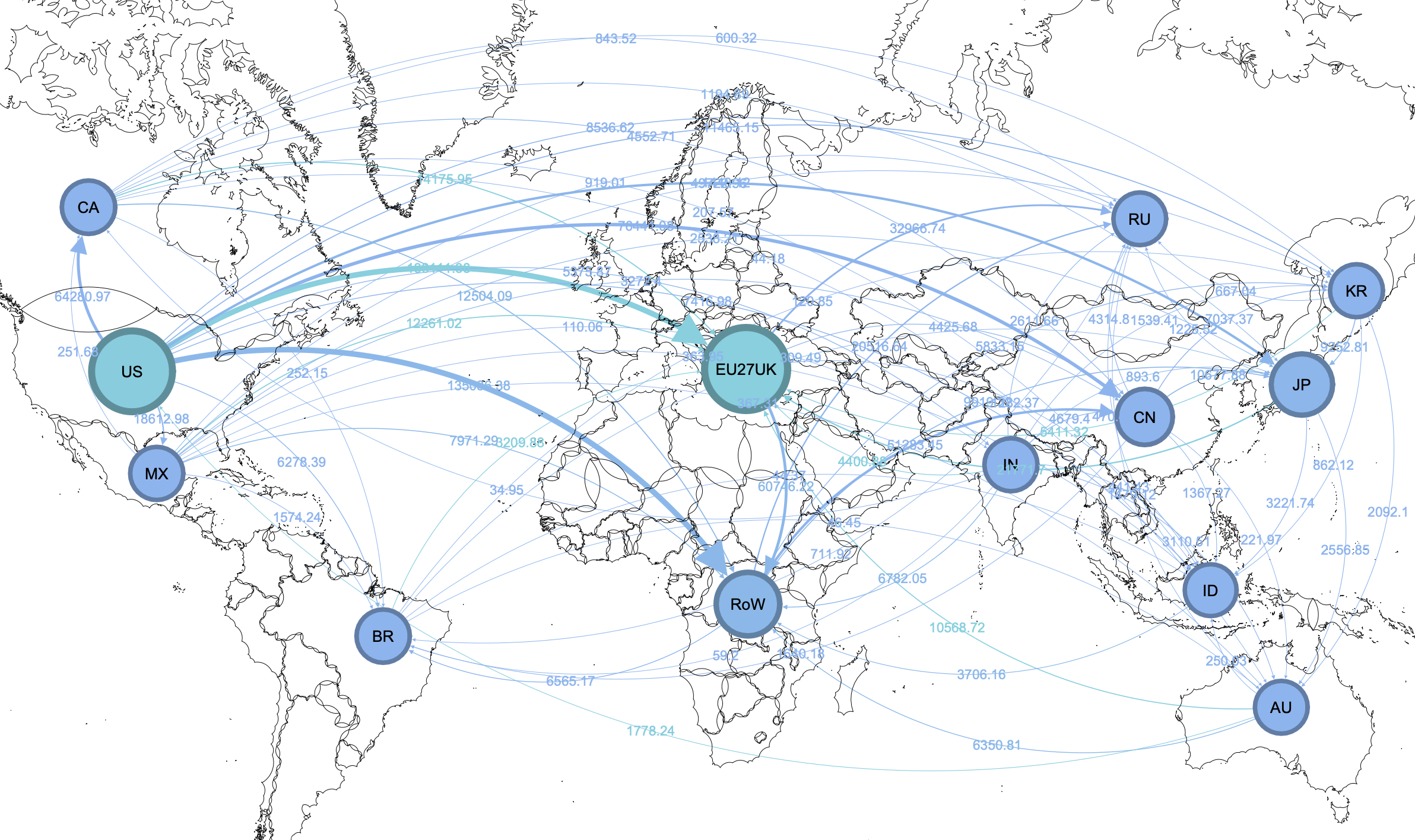}
        \caption{Value-added network in period 2}
        \label{fig:vn02}
    \end{subfigure}

    \begin{subfigure}[b]{0.49\textwidth}
        \includegraphics[width=\textwidth]{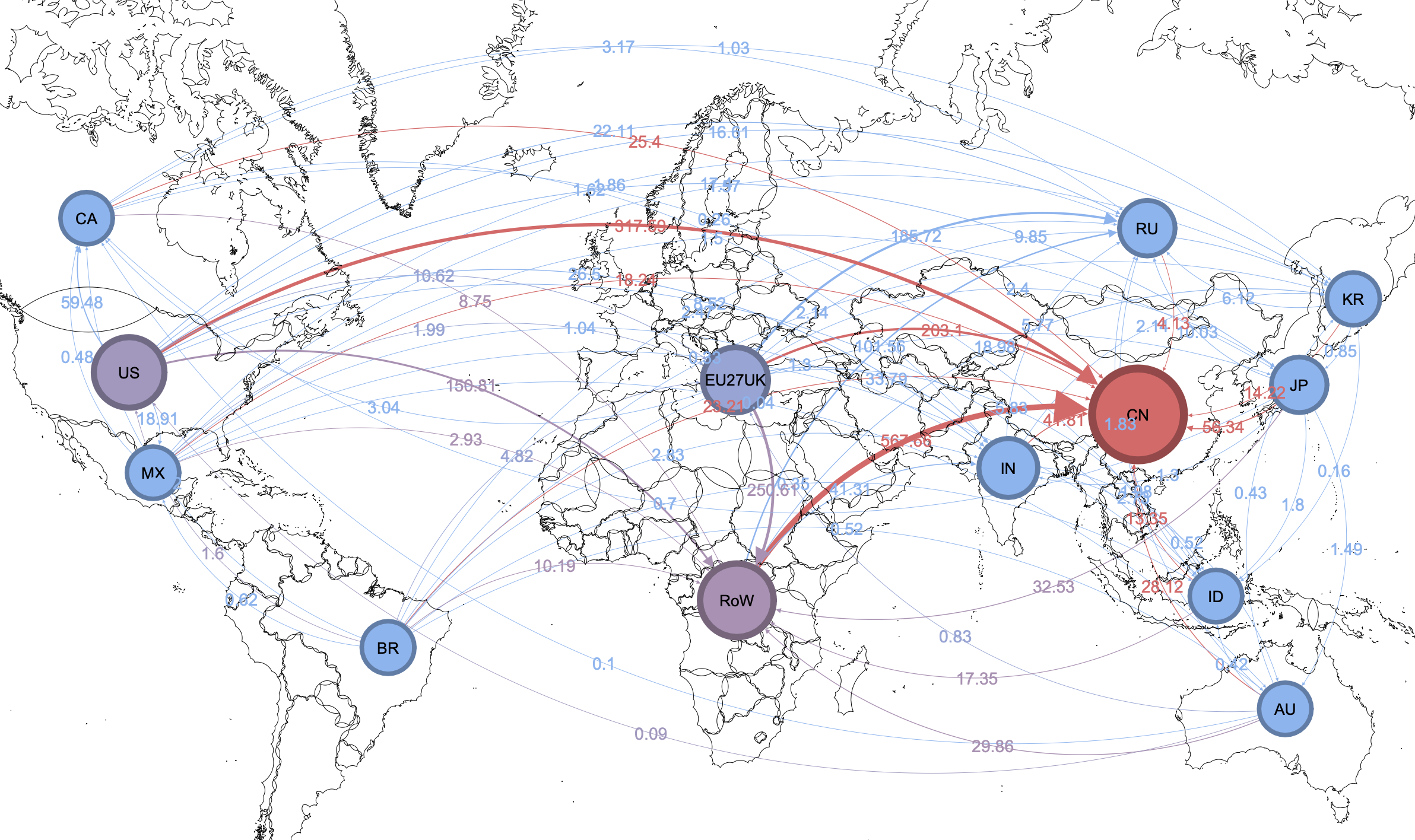}
        \caption{Emission network in period 3}
        \label{fig:en09}
    \end{subfigure}
    \begin{subfigure}[b]{0.49\textwidth}
        \includegraphics[width=\textwidth]{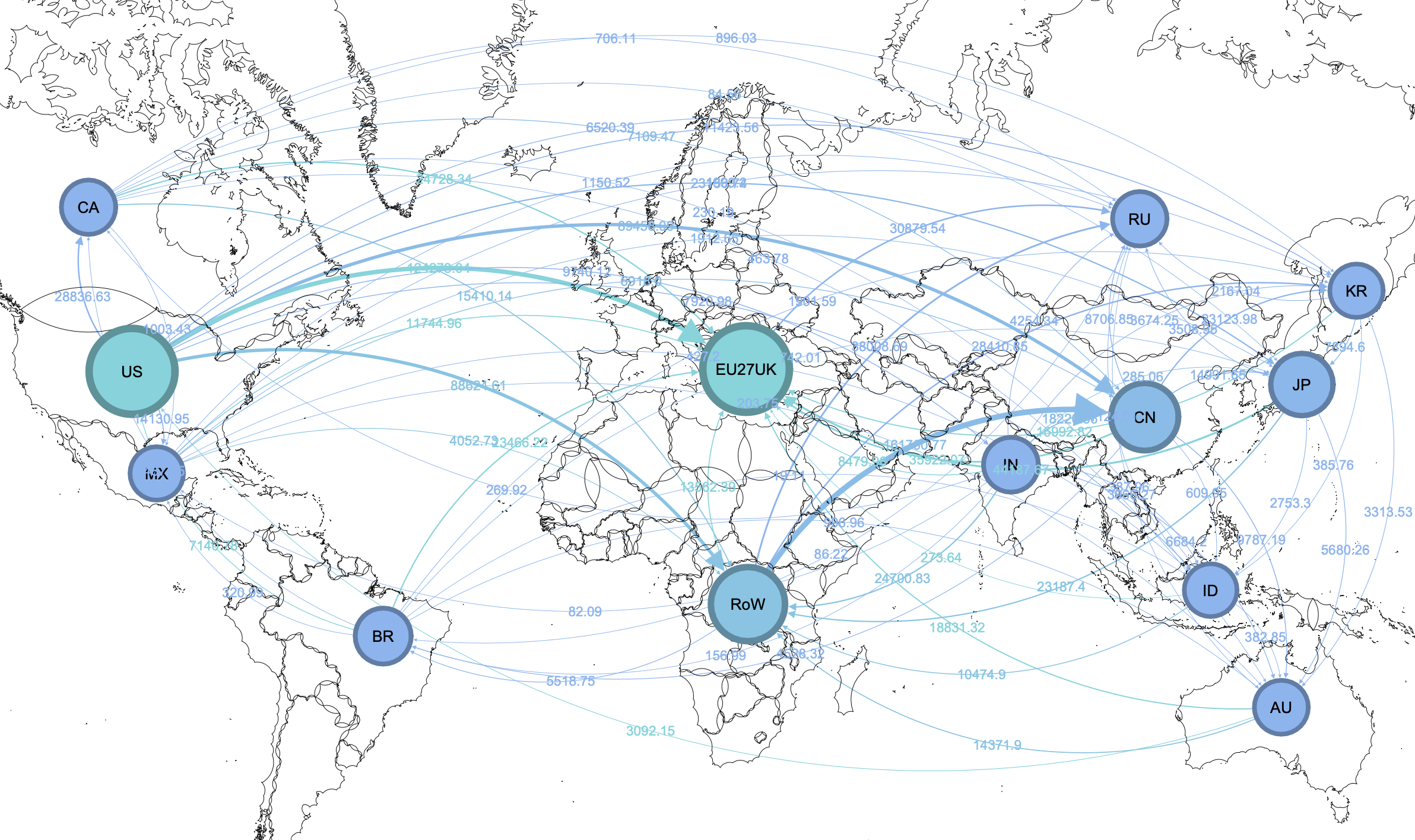}
        \caption{Value-added network in period 3}
        \label{fig:vn09}
    \end{subfigure}

    \begin{subfigure}[b]{0.49\textwidth}
        \includegraphics[width=\textwidth]{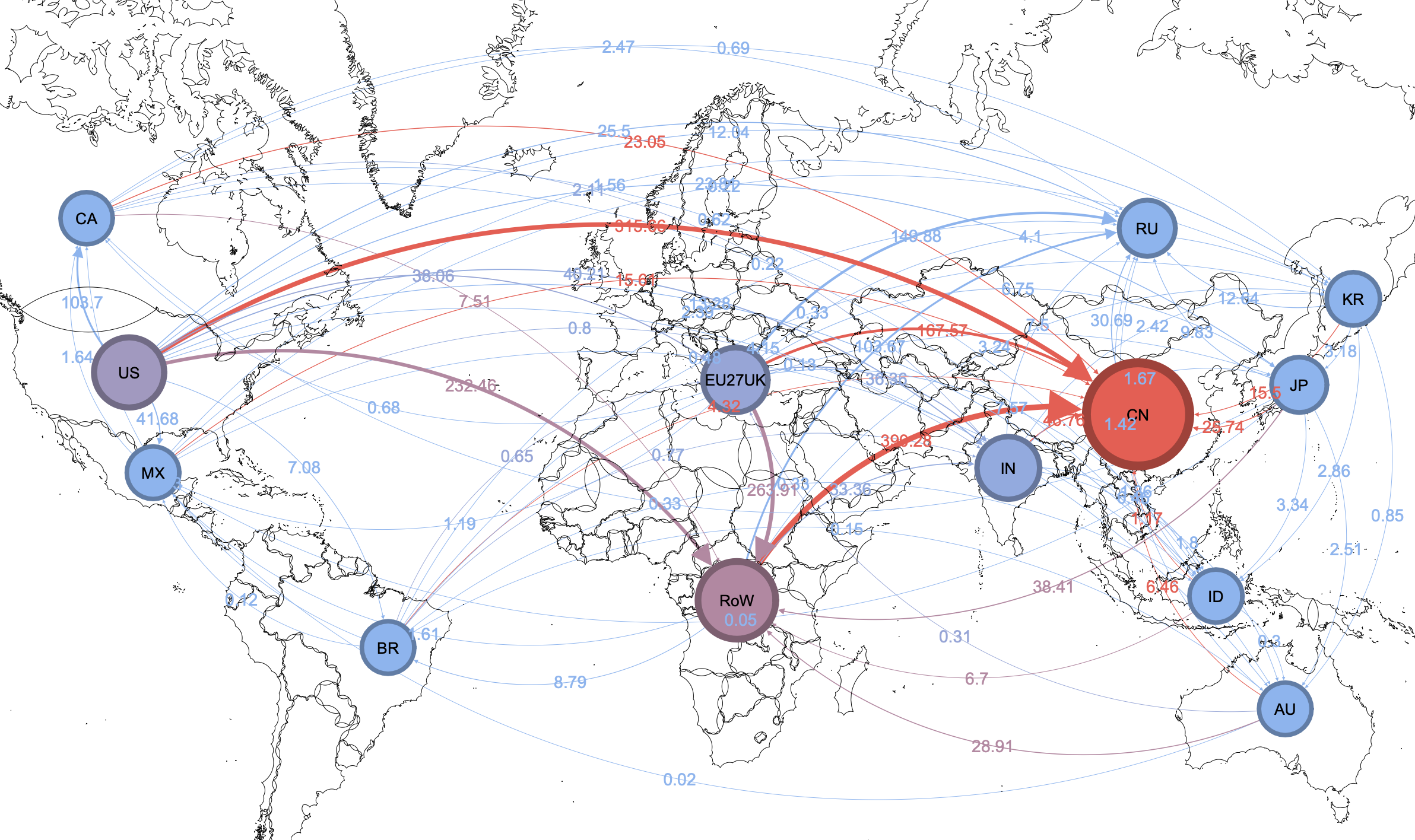}
        \caption{Emission network in period 4}
        \label{fig:en16}
    \end{subfigure}
    \begin{subfigure}[b]{0.49\textwidth}
        \includegraphics[width=\textwidth]{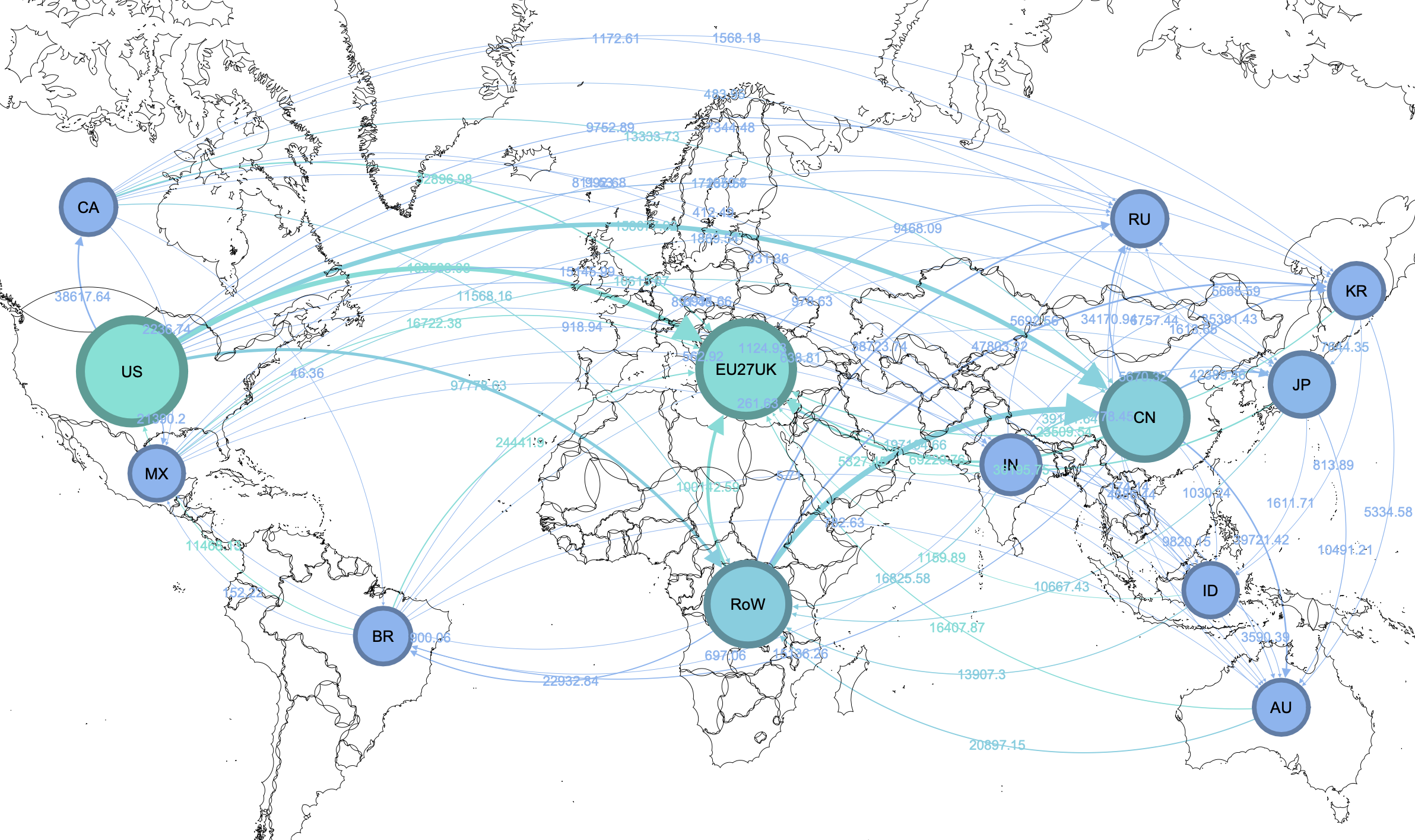}
        \caption{Value-added network in period 4}
        \label{fig:vn16}
    \end{subfigure}
    
    \caption{The dynamic networks of emission footprints (unit in million ton, Mt) and value-added footprints (unit in million Euro, M.EUR) across four distinct periods: \textbf{1) 1995 to 2001}, \textbf{2) 2002 to 2008}, \textbf{3) 2009 to 2015}, and \textbf{4) 2016 to 2022}. Nodes represent domestic emissions and value-added, while edges indicate the net flow of international footprints. These networks highlight a concerning trend: developing regions exhibit a significant increase in emissions over time but only a modest rise in value-added. This disparity in growth rates underscores a widening gap in carbon inequality, where the economic gains do not correspond proportionately with the environmental costs incurred.}
    \label{fig:gtn}
\end{figure}

By constructing dynamic global trade networks, nodes are defined as the domestic economic or emission values of countries or regions in Figure \ref{fig:gtn}. Edges are characterised by net economic and net emission flows, allowing for an intuitive understanding of the direction of emission movement and the mismatch flow capturing regional carbon inequality. This study addresses the limitations of previous studies that focused solely on single variables \cite{han2020china, jiang2019structural}, providing a comprehensive approach to better understand the underlying relationship between economic and environmental factors.

Additionally, past research has often been constrained by scale, spatial resolution, and temporal resolution, primarily focusing on national rather than international changes \cite{gao2018interprovincial, du2018network, lv2019study, zhu2022carbon, sun2020analyzing}. Our study not only emphasises the structural changes within the global trading network at an international level but also improves temporal resolution, offering insights into the structural changes in the global trade network, flow direction, the transfer of emissions, and the evolving trends of carbon inequality. These findings highlight the imbalances in trade and provide more comprehensive information for global climate decision-making.

Inequitable trade relationships can be observed by comparing the emission network with the economic network. For example, in period 1, the EU27/UK transferred 208 Mt of emission burden to the Rest of the World and 0.5 Mt to the US while still reaping economic benefits of 7K M.EUR from the Rest of the World and 43K M.EUR from the US. This imbalance has worsened over time. By period 4, the EU27/UK had transferred 263 Mt of emissions to the Rest of the World and gained 100K M.EUR in economic benefits, a tenfold increase from period 1. This demonstrates that the EU27/UK not only offloaded a substantial emission burden onto other regions through global trade but also continued to extract significant economic benefits. The recipient regions of these emissions include both developed and developing areas. This growing inequality underscores the urgent need for climate policies to address and mitigate these disparities.
\subsection{Global Carbon Inequality Network}
\begin{figure}[h!]
    \centering
    \begin{subfigure}[b]{0.49\textwidth}
        \includegraphics[width=\textwidth]{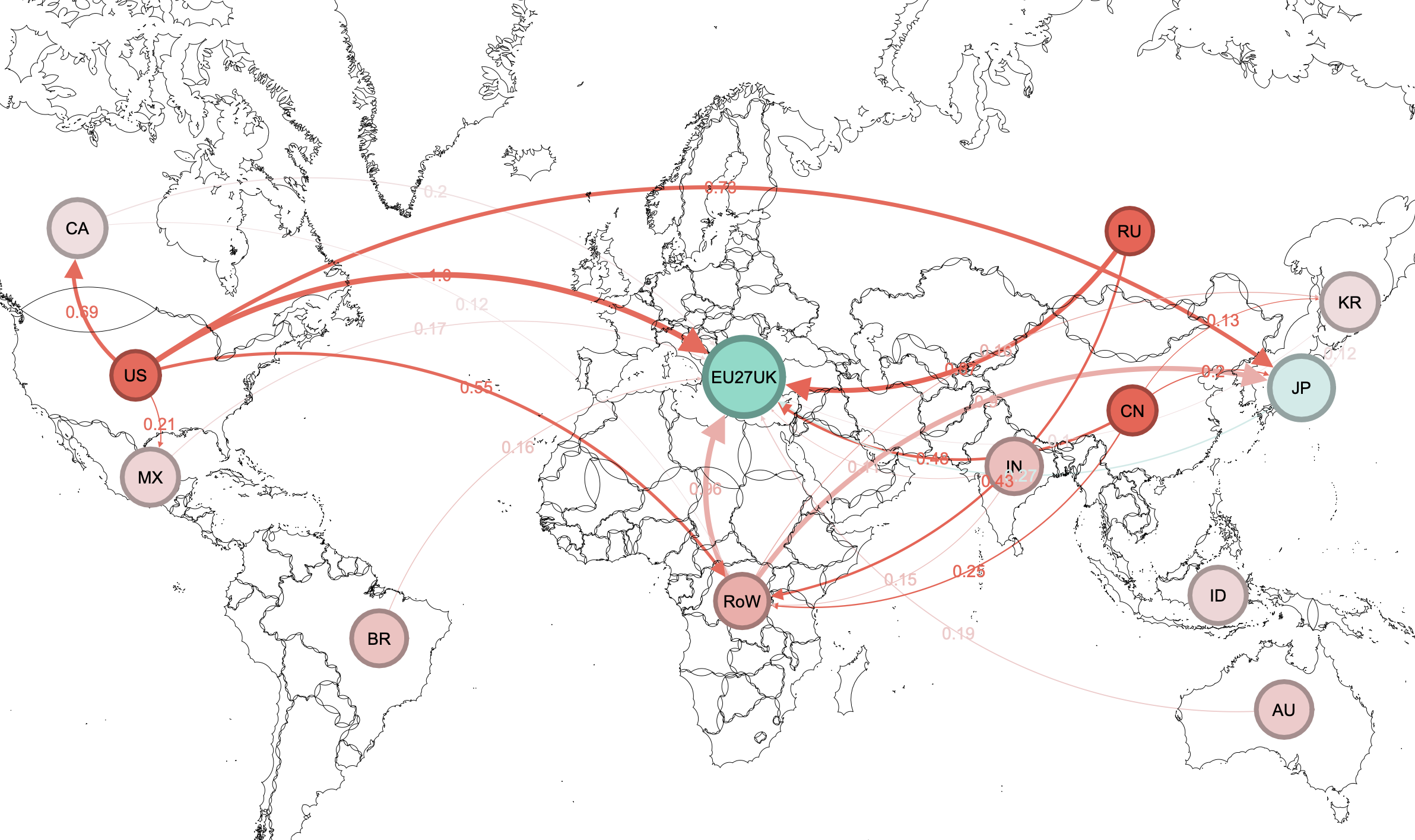}
        \caption{Carbon inequality network in period 1}
        \label{fig:eeei95}
    \end{subfigure}
    \begin{subfigure}[b]{0.49\textwidth}
        \includegraphics[width=\textwidth]{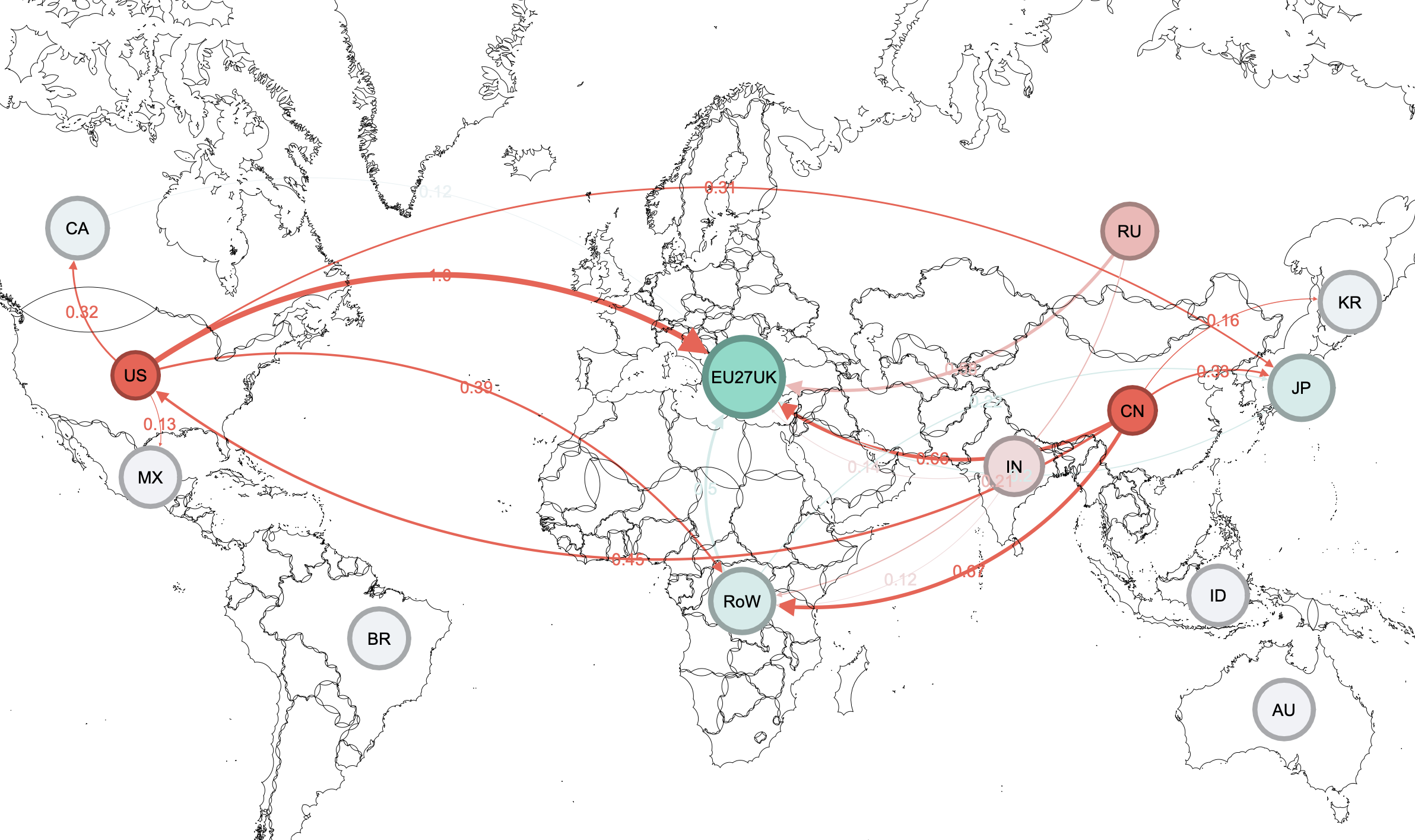}
        \caption{Carbon inequality network in period 2}
        \label{fig:eeei02}
    \end{subfigure}

    \begin{subfigure}[b]{0.49\textwidth}
        \includegraphics[width=\textwidth]{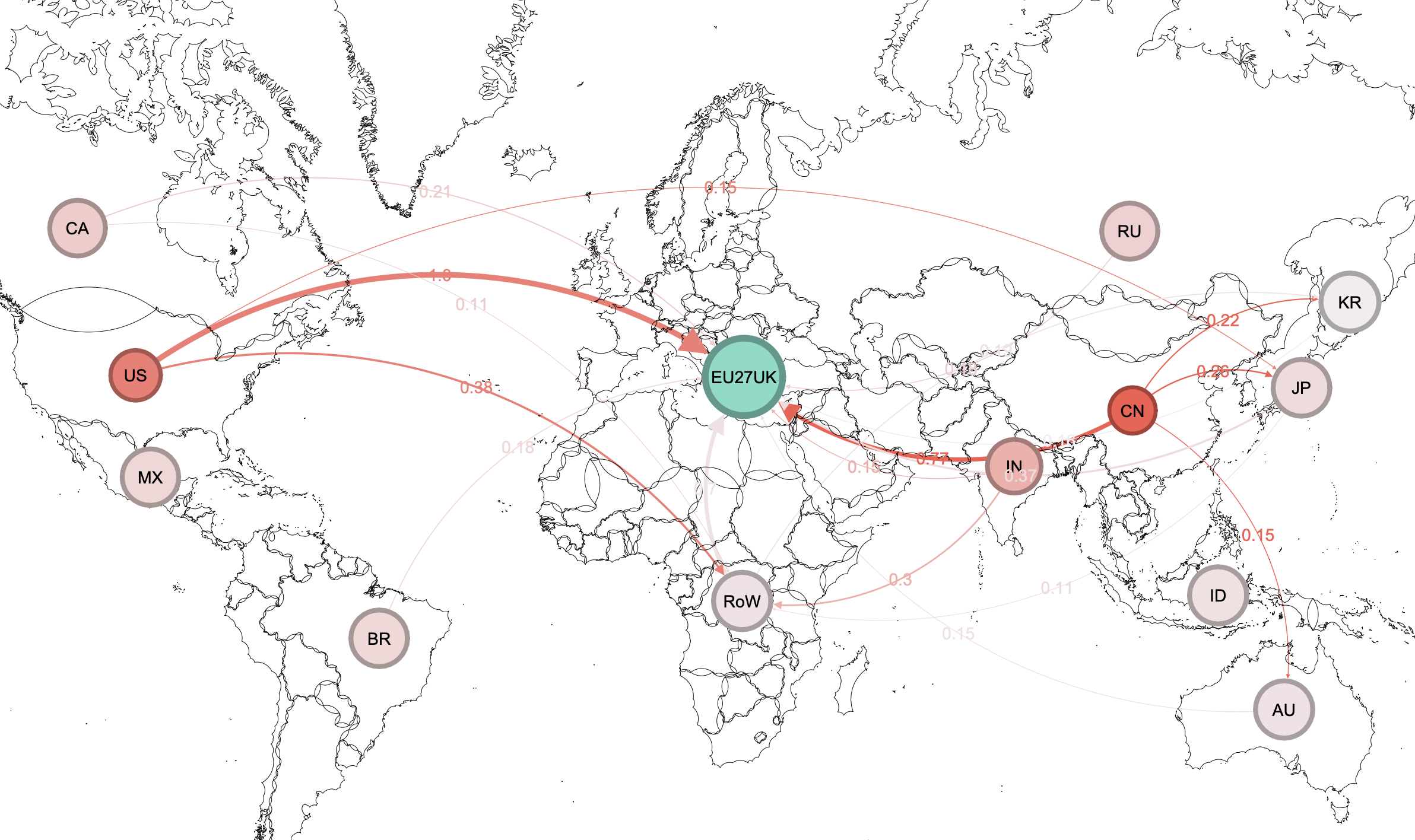}
        \caption{Carbon inequality network in period 3}
        \label{fig:eeei09}
    \end{subfigure}
    \begin{subfigure}[b]{0.49\textwidth}
        \includegraphics[width=\textwidth]{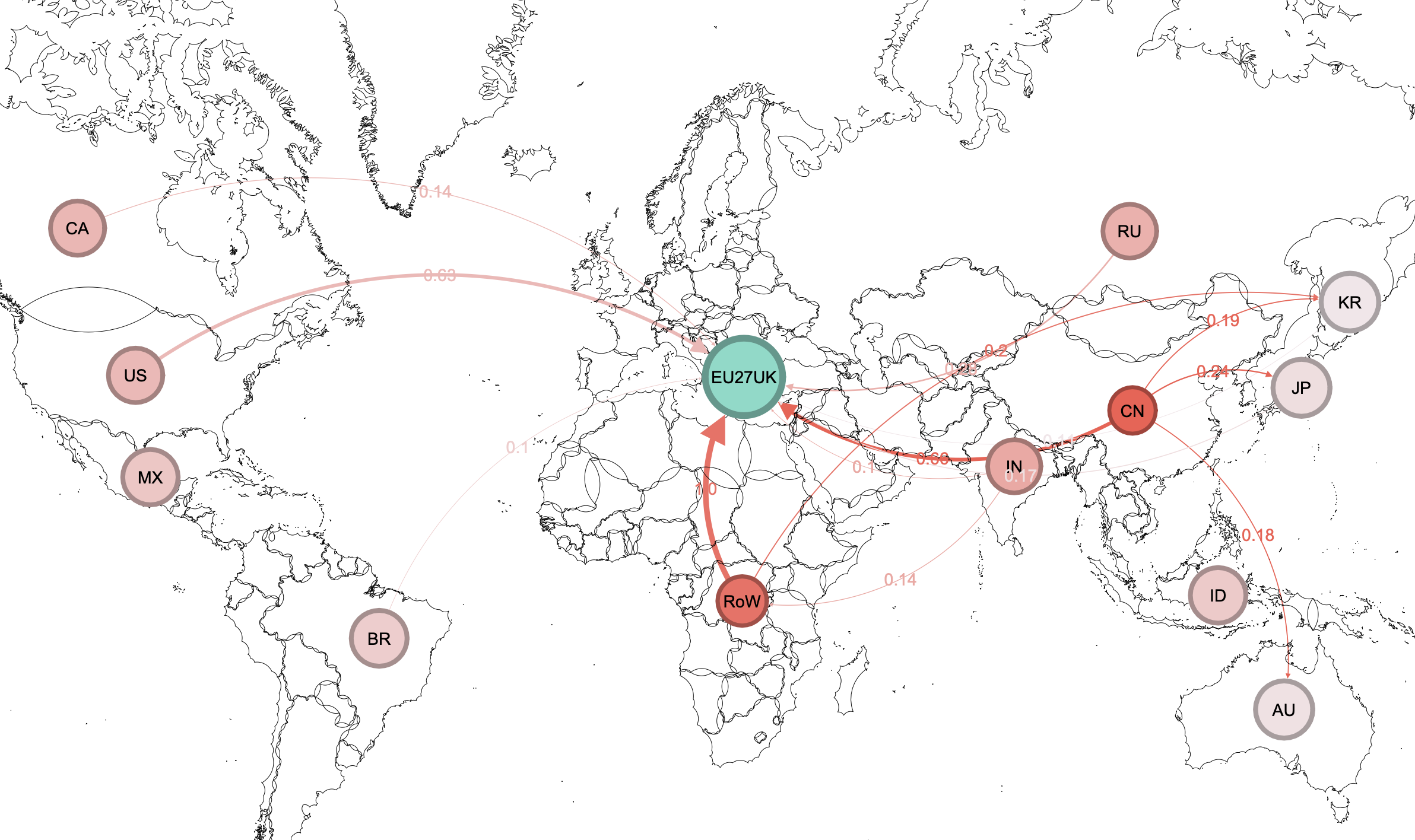}
        \caption{Carbon inequality network in period 4}
        \label{fig:eeei16}
    \end{subfigure}

    \caption{By integrating the EEEI and complex network analysis, we construct the carbon inequality network. Nodes are defined by regional EEEI values, and edges represent trades impacted by carbon inequality. The carbon inequality network highlights uneven international trades across four periods: \textbf{1) 1995 to 2001}, \textbf{2) 2002 to 2008}, \textbf{3) 2009 to 2015} and \textbf{4) 2016 to 2022}. The direction of the network indicates the regions that benefit from bilateral trade.}
    \label{fig:eeein}
\end{figure}

The global trade network, analysed through the net emission and net economic networks, reveals the existence of inequality in international trade. However, there is an inability to accurately quantify specific carbon inequality and the need to compare net emission flow and net economic flow each time to determine inequality. Our study addresses these issues by constructing a global trade carbon inequality network using the proposed EEEI metric integrated with complex network analysis. As shown in Figure \ref{fig:eeein}, nodes represent regional carbon inequality, and edges represent trade carbon inequality. These metrics range from -1 to 1, with lower values indicating more severe carbon inequality experienced by the region. In this directed graph, the direction is determined by the source and target, where the target region represents the advantageous side in bilateral trade.

A trend of increasing concentration in carbon inequality is observed, with the network showing a sparsification over time. In the early period 1, the links in the network were more dispersed, indicating carbon inequality relationships among multiple regions globally. However, by the later period 4, these links became more concentrated in a few key regions, particularly the Rest of the World (EEEI value 1), China (EEEI value 0.68), and the US (EEEI value 0.63). This indicates that the complexity of the global trade inequality network has decreased, with carbon inequality becoming more concentrated in fewer regions. This trend suggests that, during globalisation and economic integration, some regions have gradually assumed dominant positions in trade, while others bear increasing environmental burdens \cite{andrew2009approximation}. Notably, China and the Rest of the World were not primary recipients of carbon emissions in the early period. However, with rapid economic growth and its establishment as a global manufacturing hub, China, especially in period 4, has become a major recipient of carbon emissions. This phenomenon reflects China’s importance in the worldwide supply chain and exposes the growing environmental burden it faces in global trade \cite{liu2013low}. The rise of these developing countries presents significant challenges for formulating equitable climate policies.

Finding a balance in the trade-climate dilemma to reduce global carbon emissions effectively is a critical issue in sustainable development within international trade \cite{kander2015national}. We propose the EEEI-based carbon inequality network, which provides insights for decision-making that aims to minimise economic impact and enhance trade fairness while reducing overall global carbon emissions. By examining the evolution of key nodes, the study confirms that the EU27/UK has consistently maintained a favourable position in the network, not only offloading substantial carbon emissions but also reaping significant economic benefits in each period. This stable advantageous position suggests that, despite the evolving global trade network, the EU27/UK’s economic and environmental policies have been highly effective in maintaining favourable trade status. This demonstrates that the EEEI is a reliable and expandable indicator for assessing the degree of carbon inequality within the context of the global trade network.
 
\section{Discussion}
This study introduces the Ecological Economic Equality Index (EEEI) combined with complex network analysis to address the gap in quantitative measures of carbon inequality on a global scale. The results reveal significant discrepancies in net emissions and net value-added flows, showing that carbon inequality affects both developed and developing regions. Notably, the carbon inequality experienced by the US can exceed that of some developing countries, highlighting an oversight in previous studies. However, the primary victims of carbon inequality remain developing regions like China and the Rest of the World, which bear the brunt of this disparity. Despite China's significant trade surplus as a major emission exporter, it shoulders a substantial emission reduction burden, with an EEEI value of -1. In contrast, the EU27/UK benefits the most within the global trading network, with a value 1. The EEEI-based carbon inequality network identifies vital areas of unequal trade that need attention, particularly those involving the EU27/UK. Our analysis underscores the need for targeted climate policies tailored to specific regional circumstances, assessing the ecological and economic impacts of international trade.

While the EEEI is derived from net emissions and net economic benefits to measure carbon inequality, there are inherent limitations. Numerous factors influence net emissions and net economic benefits, such as carbon intensity. Regions with lower carbon intensity tend to have higher EEEI values because their export emissions are more significant than their import emissions. Consequently, the advantageous position of the EU27/UK in trade, as highlighted in this study, may also be due to its lower carbon intensity.

The utility of the EEEI extends beyond global datasets to more localised scales and is applicable to any EE-MRIO structured data, such as national-level \cite{fry2022creating}, provincial-level \cite{jiang2019provincial}, and city-level data \cite{zheng2022entropy}. This adaptability makes it a valuable tool for identifying unequal trade practices within various supply chains. In future research, we plan to conduct sector-level analyses to support sustainable investment. The open-sourced ExioNet development toolkit we provide enables future researchers to more efficiently utilise this analysis and model global trading networks. Looking forward, additional dimensions, such as social and energy impacts, could be incorporated and assessed using employment and energy usage data from the EE-MRIO emission accounting tables. This expansion would enhance the measure as a more comprehensive indicator of sustainability, further aiding efforts to mitigate global carbon inequality.

\section{Method}
\subsection{Construction of Carbon Inequality Network}
\begin{figure}[h!]
    \centering
    \includegraphics[width=1\linewidth]{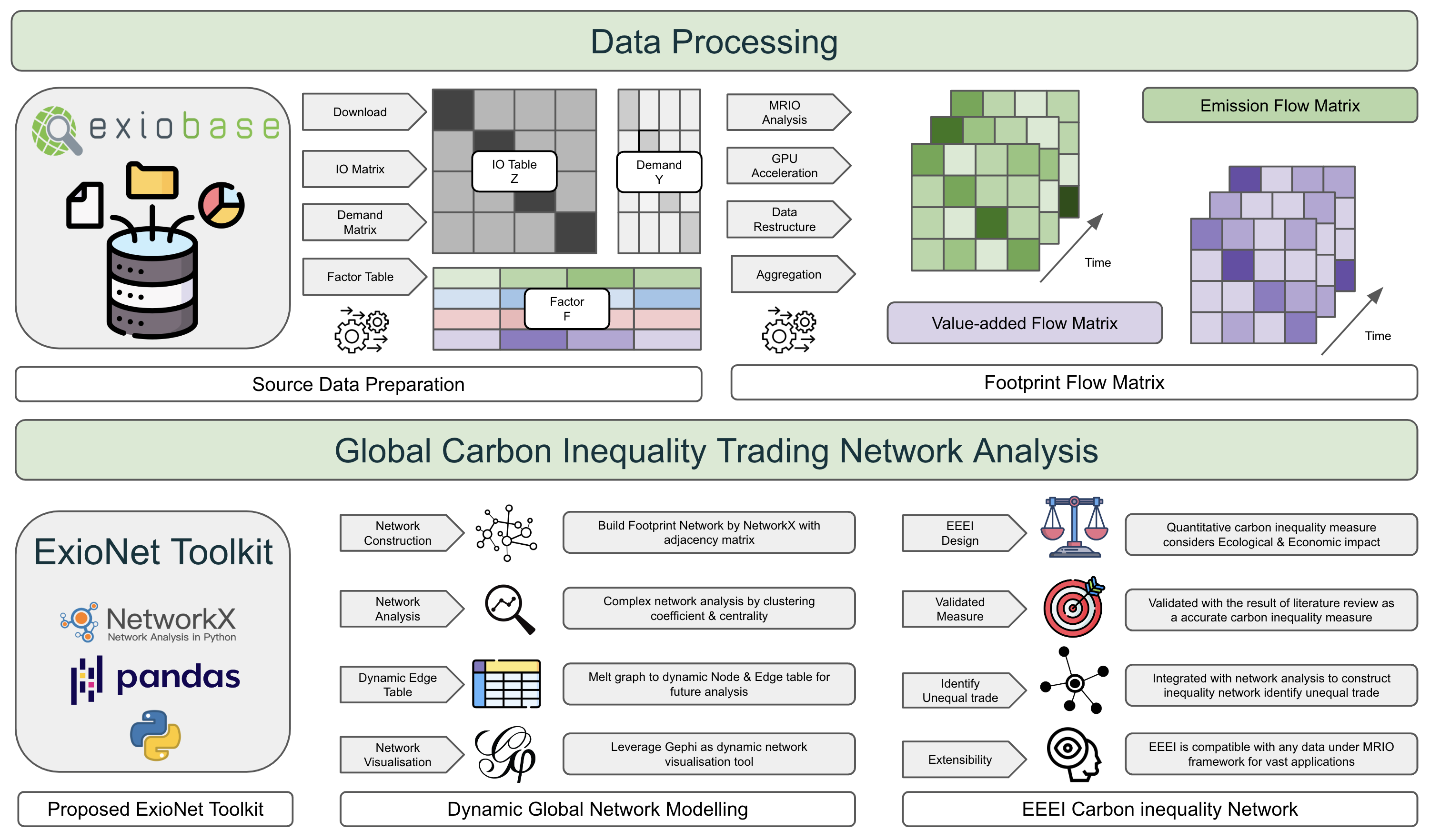}
    \caption{The pipeline of proposed ExioNet Toolkit. We calculate the footprint matrix based on the EE-MRIO ExioBase 3.8.2 framework and model the net emissions, net value-added networks, and EEEI-based carbon inequality network leverages NetworkX \cite{hagberg2008exploring} incorporating measures such as the clustering coefficient and centrality. We visualised dynamic networks in Gephi \cite{bastian2009gephi} illustrating the network's temporal structural changes.}
    \label{fig:pipeline}
\end{figure}

Figure \ref{fig:pipeline} illustrates the construction pipeline of the net emission, net value-added, and EEEI-based carbon inequality network. The calculation of the footprint matrix is based on the EE-MRIO framework, which captures inter-regional trade among \( m \) regions through the input-output matrix \( Z \). Each submatrix \( Z^{rs} \) represents the trade requirements from region \( r \) to region \( s \) for each sector. The demand matrix \( Y \) represents the final demand. The global economic output \( x \) is determined by \( Z \), \( Y \), and the summation vector \( e \):

\begin{equation}
    \begin{pmatrix}
    x^1 \\
    x^2 \\
    \vdots \\
    x^m \\
    \end{pmatrix} = \begin{pmatrix}
    Z^{11} & Z^{12} & \cdots & Z^{1m} \\
    Z^{21} & Z^{22} & \cdots & Z^{2m} \\
    \vdots  & \vdots  & \ddots & \vdots  \\
    Z^{m1} & Z^{m2} & \cdots & Z^{mm}
    \end{pmatrix} e + \begin{pmatrix}
    Y^{11} & Y^{12} & \cdots & Y^{1m} \\
    Y^{21} & Y^{22} & \cdots & Y^{2m} \\
    \vdots  & \vdots  & \ddots & \vdots  \\
    Y^{m1} & Y^{m2} & \cdots & Y^{mm}
    \end{pmatrix} e.
\end{equation}

The directed requirement matrix \( A \) is then calculated by multiplying \( Z \) with the diagonalised and inverted vector of the global economy's output \( \hat{x}^{-1} \):

\begin{equation}
    A = Z\hat{x}^{-1}.
\end{equation}

The total output \( X \) can be expressed in terms of the multiplication of the Leontief Matrix \( L \):

\begin{equation}
    X = (I - A)^{-1}y = Ly.
\end{equation}

Finally, the emission and value-added footprint matrices \( E \) and \( V \) are determined by multiplying the diagonalized emission and value-added matrices \( \hat{e} \) and \( \hat{v} \) from the emission accounting table:

\begin{equation}
    E = \hat{e}X = \hat{e}(I - A)^{-1}y.
\end{equation}

\begin{equation}
    V = \hat{v}X = \hat{v}(I - A)^{-1}y.
\end{equation}

\subsection{Ecological Economic Equality Index}
The proposed Ecological Economic Equality Index (EEEI) considers the net emission flow \( e^{net}_r \) and net value-added flow \( v^{net}_r \) of region \( r \) as the difference between total exports and total imports. Here, \( e^{rs} \) and \( v^{rs} \) indicate the export emissions and value-added flows from region \( r \) to region \( s \), respectively, while \( e^{sr} \) and \( v^{sr} \) represent the import flows. The net emission and value-added flows for region \( r \) are calculated as follows:

\begin{equation}
    e^{net}_r = \sum_{s \neq r} e^{rs} - \sum_{s \neq r} e^{sr}.
\end{equation}

\begin{equation}
    v^{net}_r = \sum_{s \neq r} v^{rs} - \sum_{s \neq r} v^{sr}.
\end{equation}

The EEEI for region \( r \) is calculated using the normalised difference between the net emission and net value-added flows:

\begin{equation}
    \text{EEEI}_r = f(f(e^{net}_r) - f(v^{net}_r)).
\end{equation}

We utilize min-max scaling normalization \( f(x) \) to scale the EEEI to the range \([-1, 1]\), defined as follows:

\begin{equation}
    f(x) = 2 \times \frac{x - x_{\min}}{x_{\max} - x_{\min}} - 1.
\end{equation}

\section{Data Availablity}\label{6}
The raw data source is ExioBase 3.8.2 hosted on Zenodo (\url{https://zenodo.org/records/5589597}). The processed data and results are available in the GitHub repository (\url{https://github.com/Yvnminc/EENet}). The sectoral-level multi-dimensional footprint networks are derived by ExioML Toolkit \cite{guo2024exioml} hosted on Zenodo (\url{https://zenodo.org/records/10604610}). 

\section{Code Availablity}\label{7}
We open-sourced the ExioNet Toolkit in Python, and all the processed data, results and visualisations are hosted in the GitHub repository (\url{https://github.com/Yvnminc/EENet}). 

\clearpage
\bibliography{sn-bibliography}

\appendix
\section{Network Analysis}

\begin{figure} [!ht]
    \centering
    \begin{subfigure}[b]{0.49\textwidth}
        \includegraphics[width=\textwidth]{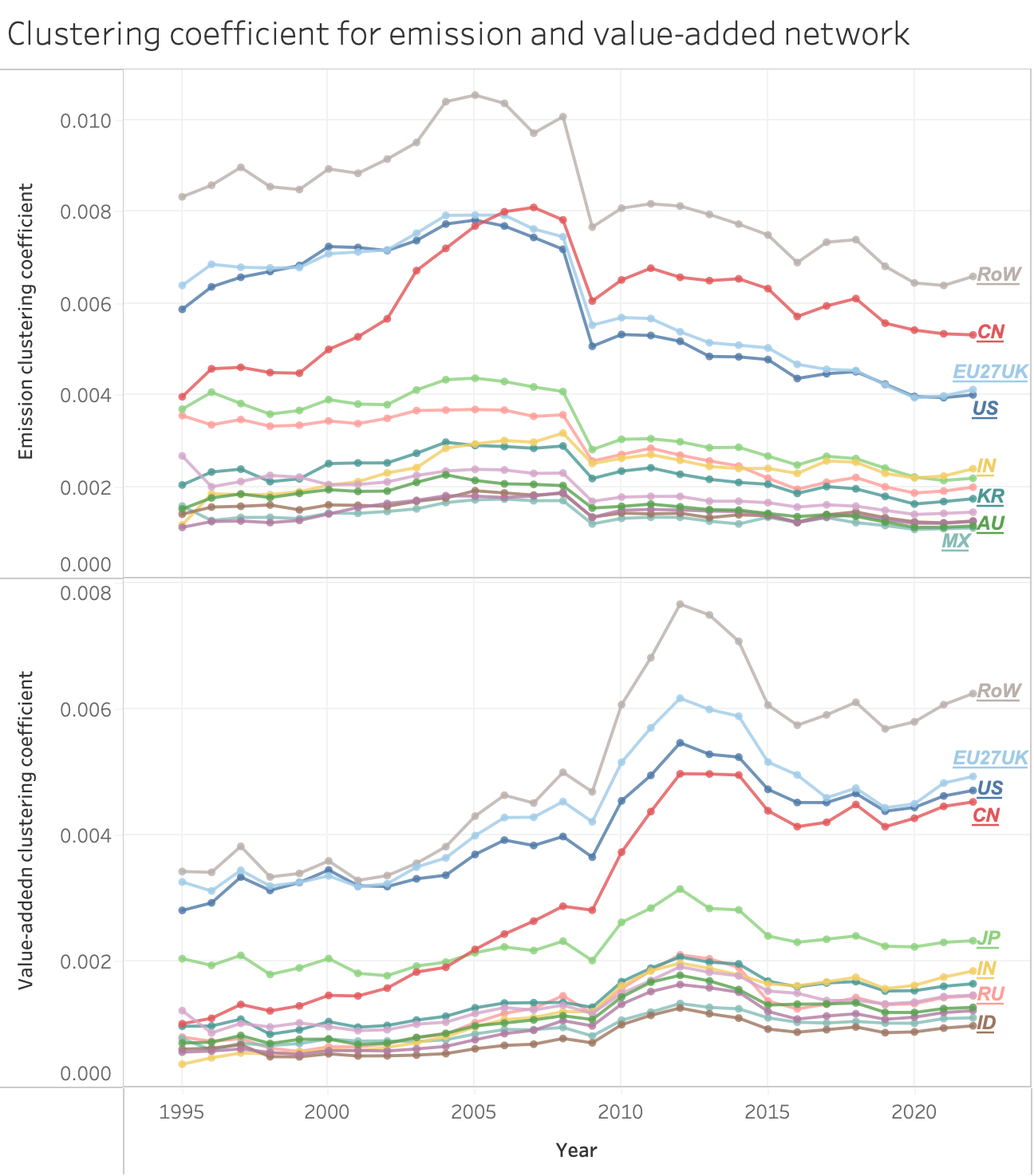}
        \caption{Clustering coefficient}
        \label{fig:cc}
    \end{subfigure}
    \begin{subfigure}[b]{0.49\textwidth}
        \includegraphics[width=\textwidth]{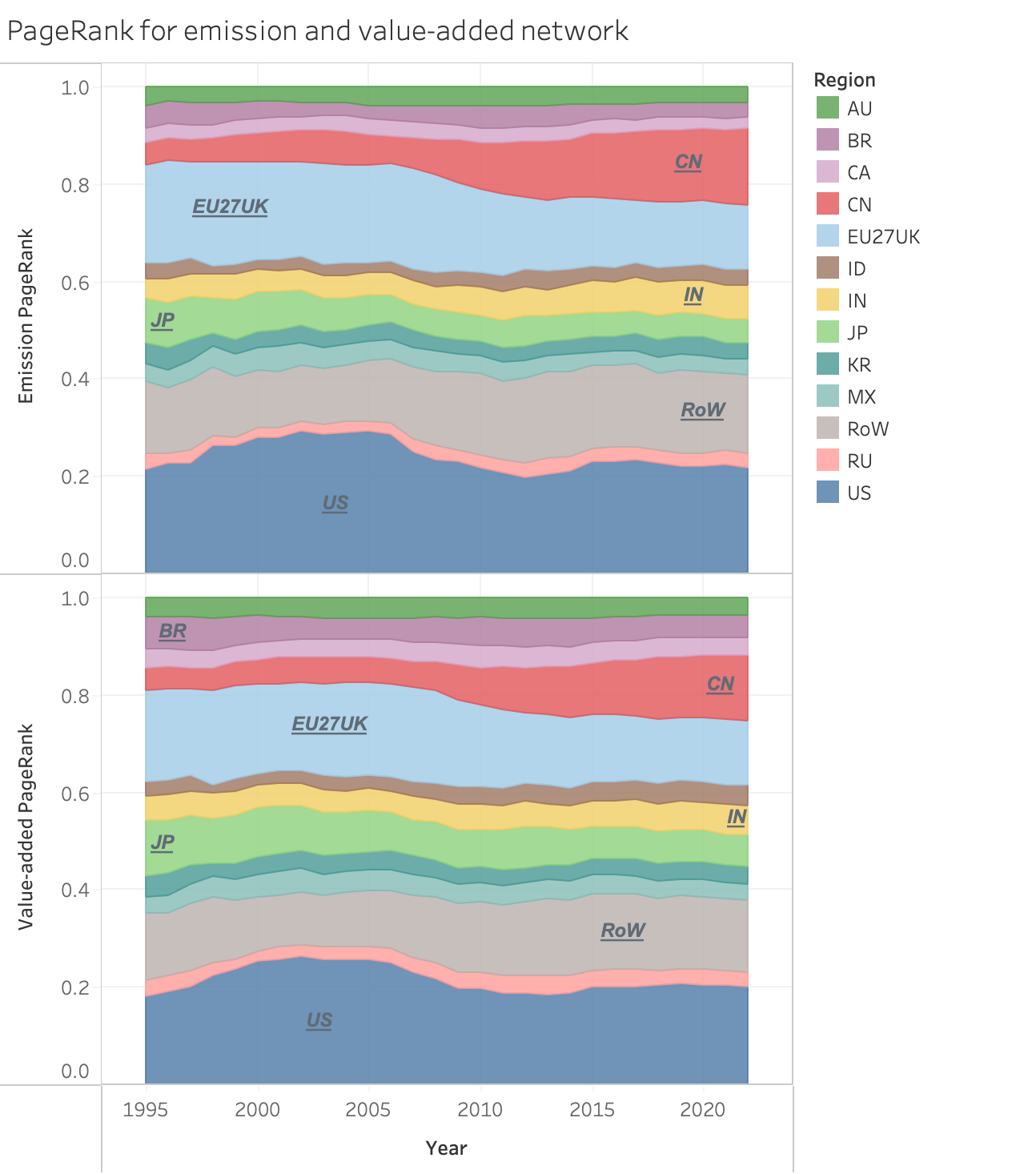}
        \caption{PageRank centrality}
        \label{fig:pg}
    \end{subfigure}
    
    \caption{Network measures of global trade networks. \textbf{Panel a)} displays the clustering coefficient, which quantifies the density of regional connections and identifies a pivotal shift around 2009, termed the 'Great Pattern Shift'. \textbf{Panel b)} illustrates the PageRank centrality, pinpointing the dominant regions in the network and highlighting China's dramatic rise in significance within the global trading network.}
    \label{fig:na}
\end{figure}

Network measures are critical for understanding the dynamics of global trade networks. The clustering coefficient and PageRank centrality provide insights into the structural changes and influential regions within these networks \cite{brin1998anatomy, onnela2005intensity, saramaki2007generalizations}.

The clustering coefficient, illustrated in Figure \ref{fig:cc}, quantifies the density of regional connections. It highlights a pivotal moment around 2009, coinciding with the global financial crisis, referred to as the "Great Pattern Shift". Before 2009, the clustering coefficients for both emissions and value-added networks showed a relatively stable trend, indicating steady trade connections and regional interdependencies. Around 2009, there was a significant drop in the clustering coefficients for both networks, reflecting a disruption in global trade connectivity due to the financial crisis. Post-crisis, the emission clustering coefficient demonstrated a gradual upward trend, suggesting a slow recovery and stabilisation of emission-related trade connections. In contrast, the value-added clustering coefficient showed a more marked increase after the crisis, indicating a quicker recovery and expansion of economic interconnections. This reflects how economic activities rebounded and even expanded beyond pre-crisis levels, particularly in regions like the EU27/UK and the US.

PageRank centrality, depicted in Figure \ref{fig:pg}, identifies key influential regions within the global trade networks. The analysis reveals that the increasing centrality of China in both the emission and value-added networks underscores its growing significance in global trade. This indicates China's pivotal role in the network's structure and dynamics, making it a critical region for influencing global trade patterns and carbon inequality. The EU27/UK and the US have consistently maintained high centrality values, reflecting their continued importance in global trade. However, their relative positions have fluctuated, highlighting shifting trade dynamics and regional influence. The RoW category shows a stable yet significant presence in the network, indicating diverse contributions from multiple regions that collectively hold substantial influence.

The network analysis using clustering coefficients and PageRank centrality provides valuable insights for policymaking. The analysis highlights the importance of targeting regions with high centrality values, such as China, EU27/UK, and the US, for climate policies. These regions have the potential to influence network-wide diffusion and optimise strategic economic and environmental outcomes. The findings emphasise the need for global cooperation and technological innovation to address carbon inequality effectively. Developed regions should support developing regions through financial and technological aid to reduce carbon emissions and enhance sustainable development. The rapid recovery of the value-added network post-crisis suggests that economic policies promoting resilience can help maintain and expand trade connections even during disruptions. Policymakers should focus on creating robust economic systems that can withstand and quickly recover from global shocks.

\section{Aggregation}
The regions included in this study and aggregation are summarised in Table \ref{table:agg}.

\begin{table} [!h]
    \centering
    \caption{Aggregation of regions included in the dataset.}
    \begin{tabular}{l||p{10cm}}
        \hline
        \textbf{Abbreviation} & \textbf{Countries or Regions} \\
        \hline
        AU & Australia \\
        \hline
        BR & Brazil \\
        \hline
        CA & Canada \\
        \hline
        CN & China \\
        \hline
        EU27/UK & Austria, Belgium, Bulgaria, Cyprus, Croatia, Czech Republic, Denmark, Estonia, Finland, France, Germany, Greece, Hungary, Ireland, Italy, Latvia, Lithuania, Luxembourg, Malta, Netherlands, Poland, Portugal, Romania, Slovakia, Slovenia, Spain, Sweden, the United Kingdom \\
        \hline
        ID & Indonesia \\
        \hline
        IN & India \\
        \hline
        JP & Japan \\
        \hline
        KR & South Korea \\
        \hline
        MX & Mexico \\
        \hline
        RoW & RoW Asia and Pacific, RoW America, RoW Europe, RoW Africa, RoW Middle East, Taiwan, Turkey, South Africa, Norway, Switzerland \\
        \hline
        RU & Russia \\
        \hline
        US & United States \\
        \hline
    \end{tabular}
    \label{table:agg}
\end{table}

\section{Competing Interests}
The authors declare no competing interests.

\section{Author Contributions}
Y. Guo and J. Ma designed the study. Y. Guo proposed the EEEI carbon inequality metric, processed the dataset, network modelling, coding, visualisations, and completed the paper draft. J. Ma supervised the project. Both authors participated in writing the paper. C. Guan participated in the project design discussion and helped improve the paper draft.

\end{document}